\documentclass[11pt]{article} \input{psfig} % \thesaurus{ } %
\usepackage{epsfig}
\title{A five dimensional model of varying effective gravitational
and fine structure constants}

\author{J.P. Mbelek and M. Lachi\`eze-Rey \\ Service d'Astrophysique,
C.E. Saclay \\ F-91191 Gif-sur-Yvette Cedex, France}

\begin{document} \maketitle \baselineskip=4mm

\begin{abstract} We explore the possibility that the reported time
variation of the fine structure constant $\alpha$ is due to a
coupling between electromagnetism and gravitation. We predict such a
coupling from a very simple {\sl effective} theory of physical
interactions, under the form of an improved version of the
Kaluza-Klein theory. We show that it precisely leads to a variation
of the effective fine structure constant with cosmic conditions, and
thus with cosmic time. The comparison with the recent data from
distant quasars absorption line spectra gives a good agreement;
moreover, this may reconcile the claimed results on $\alpha$ with the
upper limit from the Oklo naturel Uranium fission reactor.

\end{abstract}


\section{Introduction}\label{intro}

Recent publications \cite{Webb, Murphy} report observations of
distant quasars absorption lines, which may indicate a time variation
of the fine structure constant $\alpha$. Different kinds of
explications have been proposed (see in conclusion), which all
involve new physics \cite{Barrow, others}. Since many theoretical
arguments suggest that our present theories of physical interactions
are not the ultimate ones, this possibility deserves serious
attention.

An ultimate theory would include additional fields and coupling which
remain presently unknown. At the effective level, and in particular
in astrophysical conditions, those can be manifest as a soft
dependence of the \textquotedblleft constants" of the interactions
with some parameters. For instance, the effective theory considered
here predicts a variation of the effective gravitatonal constant $G$
with respect to electromagnetic conditions, and a variation of the
effective $\alpha$ with respect to gravitational conditions, and thus
with the cosmic time (seen as a parameter expressing the variation of
the cosmic gravitational potential, through the Friedmann - Lema\^\i
tre equations). In this paper, we calculate the expected variation
of~$\alpha$, and compare it with astronomical observations. In a
companion paper (\cite{paper I}), we have shown that the predicted
dependence of $G_{eff}$ with the value of the geomagnetic field, in
the same framework, may explain the discordant terrestrial
measurements of the gravitational constant.

One of the simplest effective theories that it is possible to build
(beside Brans-Dicke type theories) results from the compactification
of the Kaluza-Klein (KK) one. As we show here (see also
\cite{Lichnerowicz}), this leads to replace the gravitational
constant $G$ and the fine structure constant $\alpha$ by {\sl
effective values}, which vary with the scalar field $\Phi$ introduced
in the theory. On the other hand, different authors \cite{Witten,
Salam} have pointed out that a pure KK action leads to instability of
the theory, because of the bad sign for the kinetic term of $\Phi$
\cite{Sokolowski}. The same authors suggested the presence of an
additional field to cure this problem. In a previous study
\cite{Mbelek}, we applied an argument initially from Landau and
Lifshitz \cite{Landau}, to study this instability, and we proposed a
minimal hypothesis for stabilization: the addition of an external
field $\psi$. Thus we consider this modified KK theory (hereafter
KK$\psi$) as the simplest candidates for an effective theory, a
prototype to explore the possibility that the observational results
are due to a coupling between gravitation and electromagnetism
(hereafter GE coupling).

Seen in our 4-dimensional space-time, the KK$\psi$ Lagrangian leads
to a theory of gravitation and electromagnetism, with the additional
fields $\Phi$ (internal KK field) and $\psi$ (external stabilizing
field). The latter induce a GE coupling, which appears precisely as a
dependence of the (effective) constants $G_{eff}$ and $\alpha_{eff}$
with respect to other fields. This paper explores the cosmic
evolution of $\alpha_{eff}$ generated by that of matter and
gravitation (spacetime curvature).

In section 2, we calculate the variations predicted by the KK$\psi$
theory: section 2.1 recalls the definition of the effective coupling
constants, and gives the effective Maxwell-Einstein equations in the
context of the compactified KK theory; section 2.2 introduces the
KK$\psi$ Lagrangian and the resulting equations (Maxwell-Einstein and
scalar fields evolutions); section 2.3 considers the cosmological
solution (weak fields limit, in the matter-dominated epoch). In
section 3, we compare our calculations of the cosmological evolution
of the fine structure constant with the avalaible data from distant
quasars absorption lines. Also, we discuss the consistency of our
results especially with respect to the Oklo phenomenon. In section 4,
the similarities and differences with other
work are emphasized.

\section{Effective coupling constants} \subsection{The Kaluza-Klein
theory}

The original KK theory, after dimensional reduction, leads to the
effective action in the Jordan-Fierz frame ({\it e.g.}, see
\cite{Lichnerowicz, Thiry}) \begin{equation} \label{KK action J-F
frame} S_{KK,4} = - \,\int \sqrt{-g} \,\, [ \,\frac{c^{4}}{16\pi}
\,\frac{\Phi}{G} \,R \,\,+ \,\,\frac{1}{4} \,\,{\varepsilon}_{0}
\,{\Phi}^{3} \,F_{\alpha\beta} \,F^{\alpha\beta} \,\,+
\,\,\frac{c^{4}}{4\pi G} \,\frac{{\partial}_{\alpha} \Phi
\,{\partial}^{\alpha} \Phi}{\Phi} \,] \,d^{4}x, \end{equation} where
$A^{\alpha}$ is the potential 4-vector of the electromagnetic field,
$F_{\alpha\beta} = {\partial}_{\alpha} \,A_{\beta} \,-
\,{\partial}_{\beta} \,A_{\alpha}$ is the electromagnetic field
strength tensor, $\Phi$ the (internal) scalar field related by
${\hat{g}}_{44} = -\,{\Phi}^{2}$ to the fifteenth degree of freedom,
${\hat{g}}_{44}$, of the 5-dimensional metric and $G$ the (true)
gravitational constant. We emphasize that this is {\sl not} a theory
with minimally coupled scalar field. According to Lichnerowicz
\cite{Lichnerowicz}, the quantity $ G_{eff}:= \frac{G}{\Phi} $ of the
Einstein-Hilbert term, and the factor ${\varepsilon}_{0eff} =
{\varepsilon}_{0} \,{\Phi}^{3}$ of the Maxwell term in (\ref{KK
action J-F frame}), should be interpreted respectively as the
effective gravitational \textquotedblleft constant" and the effective
vacuum dielectric permittivity. These terms depend on the local (for
terrestrial experiments) or global (at cosmological scale) value of
$\Phi$, assumed to be positive defined. Clearly, the previous
considerations lead to an {\sl effective} fine structure constant
\begin{equation} \label{alphaeff} {\alpha}_{eff} = e^{2}/4\pi
{\varepsilon}_{0eff} ~\hbar c= \frac{\alpha}{{\Phi}^{3}},
\end{equation} to be compared to the {\sl true} fine structure
constant $\alpha := e^{2}/4\pi {\varepsilon}_{0}\,\hbar c$. It is
worth noticing that this does not involve any variation of the
electric charge, unlike the earlier suggestion of Bekenstein
\cite{Bekenstein}. Also, the velocity of light remains constant since
the value of the effective vacuum magnetic permeability,
${\mu}_{0eff} = {\mu}_{0}\,{\Phi}^{-3}$, insures
${\varepsilon}_{0eff} \,{\mu}_{0eff} = {\varepsilon}_{0} \,{\mu}_{0}$
(see \cite{Lichnerowicz}). Applying the least action principle to the
action (\ref{KK action J-F frame}) yields

\begin{itemize}

\item the generalized Einstein-Maxwell equations with the additional
source term \begin{equation} \label{effective Phi energy-momentum
tensor} T^{(\Phi)}_{\alpha\beta} = \frac{c^{4}}{8\pi G} \,(
\,{\nabla}_{\alpha}\,{\nabla}_{\beta}\,\Phi \,- \,g_{\alpha\beta}
\,{\nabla}_{\nu}\,{\nabla}^{\nu}\,\Phi \,), \end{equation} in
addition to the electromagnetic stress-energy tensor
$T^{(EM)}_{\alpha\beta}$. They identify to the usual expressions,
where $G$ and ${\varepsilon}_{0 }$ are replaced respectively by
$G_{eff}$ and ${\varepsilon}_{0eff}$.

\item the KK scalar field equation \begin{equation} \label{KK scalar
eq} {\nabla}_{\nu}\,{\nabla}^{\nu}\,\Phi = - \,\frac{4\pi G}{c^{4}}
\,{\varepsilon}_{0} \,{\Phi}^{3} \,F_{\alpha\beta}\,F^{\alpha\beta}.
\end{equation} \end{itemize}

\subsection{Stabilizing the Kaluza-Klein action}

To stabilize the KK action (\cite{Mbelek}), the simplest possibility
is to introduce an additional matter field: a real bulk scalar field
minimally coupled to gravity. After dimensional reduction, this field
appears as a scalar field $\psi$ in spacetime, with the effective
action (in the Jordan-Fierz frame) \begin{equation} \label{KK action
and Psi J-F frame} S_{4} = \,S_{KK,4} \,+ \,S_{\psi,4} = S_{KK,4} \,+
\,\frac{c^{4}}{4\pi G} \,\int \sqrt{-g} \,\,\Phi \,[ \,\frac{1}{2}
\,{\partial}_{\alpha} \psi \,\,{\partial}^{\alpha} \psi \,\,- \,\,U
\,\,- \,\,J \psi \,] \,d^{4}x, \end{equation} where $U = U(\psi)$
denotes the self-interaction potential of $\psi$ and $J$ its source
term. The latter includes contributions from the matter and from
$\Phi$, both proportional to the trace of their energy-momentum
tensor, {\it viz.} $\frac{8\pi G}
{3c^{4}}\,g\,(\psi\,,\,\Phi)\,T^{\alpha}_{\alpha}$, and that of the
(traceless) electromagnetic field,
${\varepsilon}_{0}\,f(\psi\,,\,\Phi)\,F_{\alpha\beta}
\,F^{\alpha\beta}$. Generally speaking these coupling functions are
temperature dependent, with magnitude decreasing as the temperature
increases (this prevents from any significant modification of the big
bang nucleosynthesis). The necessity to recover the usual physics
whenever the $\psi$-field is not excited requires $g(v\,,\,1) = f(v\,,\,1) = 0$
and $U(v) = 0$. We have written $v$ the vacuum expectation value
(VEV) of $\psi$, such that $\frac{\partial U}{\partial \psi} (v) = 0$
(definition of the VEV of $\psi$). The definition of $G_{eff}$
implies that the VEV of~$\Phi$ is 1.

Applying the variational principle to (\ref{KK action and Psi J-F
frame}) yields \begin{equation} \label{ext. scalar field eq}
{\nabla}_{\nu} {\nabla}^{\nu} {\psi} = - \,\,J \,\,-
\,\,\frac{\partial J} {\partial {\psi}} \,\psi \,\,- \,\frac{\partial
U} {\partial {\psi}} \end{equation} and \begin{equation}
\label{stabilized KK scalar eq} {\nabla}_{\nu}\,{\nabla}^{\nu}\,\Phi
= - \,\frac{4\pi G}{c^{4}} \,{\varepsilon}_{0}
\,F_{\alpha\beta}\,F^{\alpha\beta} \,{\Phi}^{3} \,\,+ \,\,U\,\Phi
\,\,+ \,\,J \psi\,\Phi \,\,+ \,\,\frac{\partial J} {\partial {\Phi}}
\,{\Phi}^{2} \,\psi \,\,- \,\,\frac{1}{2} \,( \,{\partial}_{\alpha}
\psi \,\,{\partial}^{\alpha} \psi \,) \,\Phi. \end{equation} The
effective Einstein equations are unchanged, apart from a new source
term: the effective energy momentum tensor of $\psi$,
\begin{equation} \label{effective psi energy-momentum tensor}
T^{(\psi)}_{\alpha\beta} = \frac{c^{4}}{4\pi G_{eff}} \,[
\,{\partial}_{\alpha}\,\psi \,{\partial}_{\beta}\,\psi \,\,- \,\,(
\,\frac{1}{2} \,{\partial}_{\gamma}\,\psi \,{\partial}^{\gamma}\,\psi
\,\,- \,\,U \,\,- \,\,J \psi \,) \,g_{\alpha\beta} \,].
\end{equation} Since we know that, for the effects examined here, the
effective values are close to the usual one, we linearize the two
scalar fields around their respective VEVs. Hence, equations
(\ref{ext. scalar field eq}, \ref{stabilized KK scalar eq}) above
reduce to \begin{equation} \label{ext. scalar field reduced eq}
{\nabla}_{\nu} {\nabla}^{\nu} {\psi} = - \,\,\frac{\partial J}
{\partial {\psi}} \,v \end{equation} and \begin{equation}
\label{stabilized KK scalar reduced eq}
{\nabla}_{\nu}\,{\nabla}^{\nu}\,\Phi = - \,\frac{4\pi G}{c^{4}}
\,{\varepsilon}_{0} \,F_{\alpha\beta}\,F^{\alpha\beta} \,\,+
\,\,\frac{\partial J} {\partial {\Phi}}\,v \,\,- \,\,\frac{1}{2}
\,{\partial}_{\alpha} \psi \,\,{\partial}^{\alpha} \psi.
\end{equation}

\subsection{Cosmological solutions}

For cosmology, we assume spatially constant values of the fields and
we follow their evolutions with respect to the cosmic time, $t$.
Hence, $\psi = \psi (t)$ and $\Phi = \Phi (t)$. Besides,
$F_{\alpha\beta}\,F^{\alpha\beta}$ vanishes for the pure EM cosmic
background radiation. Thus, the cosmological equations reduce to the
effective Friedmann equation (with a cosmological constant $\Lambda$)
\begin{equation} \label{eff Friedmann equation 1} (
\,\frac{\dot{a}}{a} \,)^{2} = \frac{8\pi G}{3} \,\rho \,-
\,\frac{k\,c^{2}}{a^{2}} \,+ \,\frac{\Lambda\,c^{2}}{3} \,+
\,\frac{1}{3} \,{\dot{\psi}}^{2} \,- \,\frac{1}{6} \,( \,\ddot{\Phi}
\,+ \,6H \,\dot{\Phi} \,), \end{equation} \begin{equation} \label{eff
Friedmann equation 2} \frac{\ddot{a}}{a} = \,- \,\frac{4\pi G}{3} \,(
\,\rho \,+ \,3\,\frac{P}{c^{2}} \,) \,+ \,\frac{\Lambda\,c^{2}}{3}
\,- \,\frac{1}{3} \,{\dot{\psi}}^{2} \,- \,\frac{1}{6} \,(
\,\ddot{\Phi} \,+ \,3H \,\dot{\Phi} \,) \end{equation} and, for the
scalar fields, \begin{equation} \label{eq psi J-F cosmological rho
approx 1} \ddot{\psi} \,+ \,3H \,\dot{\psi} = \,- \,\frac{8\pi G}
{3}\,{\beta}_{\psi}\,v\,( \,\rho \,- \,3\,\frac{P}{c^{2}} \,)
\end{equation} \begin{equation} \label{eq Phi J-F cosmological rho
approx 1} \ddot{\Phi} \,+ \,3H \,\dot{\Phi} = - \,\frac{1}{2}
\,{\dot{\psi}}^{2} \,\,+ \,\,\frac{8\pi G} {3}\,{\beta}_{\Phi}\,v \,(
\,\rho \,- \,3\,\frac{P}{c^{2}} \,). \end{equation} The dot denotes
the derivative with respect to the cosmic time, $H = \dot{a}/a$ is
the expansion rate (Hubble parameter), $a = a(t)$ the scale factor,
$k$ is the spatial curvature parameter and~$P$ the pressure; we have
set ${\beta}_{\psi} = \frac{\partial g}{\partial
\psi}\,(v\,,\,1)$ and ${\beta}_{\Phi} = \frac{\partial g}{\partial \Phi}\,(v\,,\,1)$. The smallness of the
observed effects implies $\mid {\beta}_{\Phi}\,v \mid \,\ll 1$, $\mid
{\beta}_{\psi}\,v \mid \,\ll 1$, whereas the consistency of the model
implies $\mid \dot{\psi} \mid \,\ll H$ and $\mid \dot{\Phi} \mid
\,\ll H$ (all confirmed by the numerical calculations below). Hence,
the small excitations of the scalar fields do not modify
significantly the variation of the scale factor with respect to the
cosmic time. As this is suggested by observations, we assume zero
spatial curvature. Let us emphasize that
equation (\ref{eq Phi J-F cosmological rho approx 1}) implies that
the extrema of $\Phi$ are necessary maxima during the matter or
matter-$\Lambda$ dominated era (present era). On account of equation
(\ref{eq psi J-F cosmological rho approx 1}), the same conclusion
applies to $\psi$ under the condition ${\beta}_{\psi}\,v > 0$
(choosing ${\beta}_{\psi}\,v < 0$ would lead to minima of $\psi$,
instead).

\subsubsection{Radiation era} Before the recombination, the content
of the universe is well described by the equation of state $P =
\frac{1}{3} \,\rho \,c^{2}$ (matter negligible, no spatial 
curvature): putting $H =
1/2t$, $a = a(t_{0}) \,(t/t_{0})^{1/2}$, $\rho = \rho (t_{0})
\,(t_{0}/t)^{2}$, we obtain \begin{equation} \label{eq psi J-F
cosmological rho approx 1 radiation era} \ddot{\psi} \,+
\,\frac{3}{2t} \,\dot{\psi} = 0, \end{equation} \begin{equation}
\label{eq Phi J-F cosmological rho approx 1 radiation era}
\ddot{\Phi} \,+ \,\frac{3}{2t} \,\dot{\Phi} = - \,\frac{1}{2}
\,{\dot{\psi}}^{2}, \end{equation} where $t_{0}$ is the present time.
The solutions of equations (\ref{eq psi J-F cosmological rho approx 1
radiation era}) and (\ref{eq Phi J-F cosmological rho approx 1
radiation era}) take the forms \begin{equation} \label{sol. psi J-F
cosmological rho approx 1 radiation era} \psi = v \,+ \,\delta\psi
(t_{d}) \,\,( \,\frac{t_{d}}{t} \,)^{1/2} \end{equation} and
\begin{equation} \label{sol. eq Phi J-F cosmological rho approx 1
radiation era} \Phi = 1 \,+ \,\frac{1}{4} \,\delta\psi (t_{d})^{2}
\,\,\frac{t_{d}}{t} \,+ \,[ \,\delta\Phi (t_{d}) \,- \,\frac{1}{4}
\,\delta\psi (t_{d})^{2} \,] \,\,( \,\frac{t_{d}}{t} \,)^{1/2},
\end{equation} where $\delta\psi (t_{d}) = \psi (t_{d}) \,- \,v$,
$\delta\Phi (t_{d}) = \Phi (t_{d}) \,- \,1$ and $t_{d}$ denotes the
epoch of matter-radiation decoupling. Now, requiring that both $\psi$
and $\Phi$ be bounded at any time past the big bang involves
$\delta\psi (t_{d}) = 0$ and $\delta\Phi (t_{d}) = 0$. Hence, both
scalar fields remain constant and equal to their respective VEV,
during the radiation era. As a consequence, the effective fine
structure constant identifies to the true fine structure constant
during the radiation era.

\subsubsection{A model without cosmological constant}

After recombination, the matter and the cosmological constant play
their role. We explore two different cosmological models:  first, in
this section,
an Einstein - de  Sitter model, with no cosmological constant and the critical
density (no spatial curvature, no pressure). In the next section we
explore a more realistic model with a cosmological constant.  Putting
$H = 2/3t$, $a = a(t_{0}) \,(t/t_{0})^{2/3}$, $\rho = \rho (t_{0})
\,(t_{0}/t)^{2}$, we obtain \begin{equation} \label{eq psi J-F
cosmological rho approx 1 bis} \ddot{\psi} \,+ \,\frac{2}{t}
\,\dot{\psi} = \,- \,\frac{4}{9} \,\frac{{\beta}_{\psi}~v}{t^{2}},
\end{equation} \begin{equation} \label{eq Phi J-F cosmological rho
approx 1 bis} \ddot{\Phi} \,+ \,\frac{2}{t} \,\dot{\Phi} = -
\,\frac{1}{2} \,{\dot{\psi}}^{2} \,\,+ \,\,\frac{4}{9}
\,\frac{{\beta}_{\Phi} ~v}{t^{2}}. \end{equation} The solutions of
equations (\ref{eq psi J-F cosmological rho approx 1 bis}) and
(\ref{eq Phi J-F cosmological rho approx 1 bis}) take the forms
\begin{equation} \label{sol. psi J-F cosmological rho approx 1 bis}
\psi = v \,+ \,\delta\psi_{0} \,\,\frac{t_{0}}{t} \,-
\,\frac{4}{9}\,{\beta}_{\psi}\,v\,\ln( \,\frac{t}{t_{0}} \,)
\end{equation} and \begin{equation} \label{sol. eq Phi J-F
cosmological rho approx 1} \Phi = 1 \,+ \,( \,\delta\Phi_{0} \,+
\,\frac{1}{4} \,\delta\psi_{0}^{2} \,) \,\frac{t_{0}}{t} \,-
\,\frac{1}{4} \,\delta\psi_{0}^{2} \,\,( \,\frac{t_{0}}{t} \,)^{2}
\,+ \,\frac{4}{9}\,v\,[ \,{\beta}_{\Phi} \,+
\,{\beta}_{\psi}\,\delta\psi_{0} \,\,\frac{t_{0}}{t} \,] \,\ln(
\,\frac{t}{t_{0}} \,), \end{equation} where $t_{0}$ denotes the
present epoch in the cosmic time, and we have set $\delta\psi_{0} =
\psi (t_{0}) \,- \,v$ and $\delta\Phi_{0} = \Phi (t_{0}) \,- \,1$. On
account of the results obtained previously at the epoch of
matter-radiation decoupling, on gets in addition the following
constraints \begin{equation} \label{constraint at matter-radiation
decoupling epoch psi} \delta\psi_{0} = -
\,\frac{4}{9}\,{\beta}_{\psi}\,v \,\frac{t_{d}}{t_{0}} \,\ln(
\,\frac{t_{0}}{t_{d}} \,) \end{equation} and \begin{equation}
\label{constraint at matter-radiation decoupling epoch Phi}
\delta\Phi_{0} = \frac{1}{4} \,\delta\psi_{0}^{2} \,\,\frac{t_{0} \,-
\,t_{d}}{t_{d}} \,+ \,\frac{4}{9}\,v\,[ \,{\beta}_{\Phi}
\,\frac{t_{d}}{t_{0}} \,+ \,{\beta}_{\psi}\,\delta\psi_{0} \,] \,\ln(
\,\frac{t_{0}}{t_{d}} \,). \end{equation}

\subsubsection{Model with cosmological constant}

\begin{enumerate}

\item {\bf Present era}

The present cosmological data seem to favor a model where $\Omega
_{\Lambda}=2\Omega _{m}\approx 2/3$, that we explore now (still no
spatial curvature and no pressure). We write the solution as $H =
H_{0} \,\sqrt{\lambda} \,\coth{x}$, $a = a(t_{0})
\,(\Omega_{m}/\lambda)^{1/3} \,\,[\sinh{x}]^{2/3}$, $\rho = \rho
(t_{0}) \,[a(t_{0})/a(t)]^{3}$ where $\lambda = \Omega _{\Lambda}=
\Lambda \,c^{2}/3 \,H_{0}^{2}$, and $x = \frac{3}{2} \,\sqrt{\lambda}
\,H_{0} \,\,t$. Solving equations (\ref{eq psi J-F cosmological rho
approx 1}) and (\ref{eq Phi J-F cosmological rho approx 1}), we
obtain $$\psi = v \,+ \,\frac{2}{9}\,{\beta}_{\psi}\,v\,\ln{\lambda}
\,+ \,\delta\psi_{\Lambda} \,\sqrt{\lambda} \,\coth{x}$$
\begin{equation} \label{sol. psi J-F cosmological rho + lambda approx
1} - \,\frac{4}{9}\,{\beta}_{\psi}\,v\,[ \,1 \,- \,x \,\coth{x} \,+
\,\ln{(\sinh{x})} \,] \end{equation} and $$ \Phi = 1 \,-
\,\frac{2}{9}\,{\beta}_{\Phi}\,v\,\ln{\lambda} \,+ \,\sqrt{\lambda}
\,\,( \,\delta\Phi_{\Lambda} \,+ \,\frac{4}{9} \,{\beta}_{\psi}\,v
\,\delta\psi_{\Lambda} \,\sqrt{\lambda}$$ $$+ \,\frac{1}{4}
\,\delta\psi_{\Lambda}^{2} \,) \,\coth{x} \,+ \,\frac{4}{9}
\,{\beta}_{\Phi}\,v \,[ \,1 \,-\,x \,\coth{x} \,+ \,\ln{(\sinh{x})}
\,] $$ $$- \,\frac{1}{4} \,\delta\psi_{\Lambda}^{2} \,\lambda
\,\,\sinh^{-2}{x} \,-\,\frac{2}{9} \,{\beta}_{\psi}\,v
\,\delta\psi_{\Lambda} \,\sqrt{\lambda} \,[ \,\frac{3}{2} \,x $$ $$+
\,\frac{1}{2} \,x \,\sinh^{-2}{x} \,+ \,\frac{1}{2} \,\coth{x} \,-
\,\coth{x} \,\ln{(\sinh{x})} $$ \begin{equation} \label{sol. eq Phi
J-F cosmological rho + lambda approx 1} \,+ \,x\,\sinh^{2}{x} \,-
\,\frac{1}{4} \,\sinh{2x} \,- \,\frac{1 \,+
\,\ln(\sinh{x})}{\sinh{x}} \,]. \end{equation} The parameters
$\delta\psi_{\Lambda}$ and $\delta\Phi_{\Lambda}$ have been
introduced in such a way that, formally, they reduce respectively to
$\delta\psi_{0} = \psi (t_{0}) \,- \,v$ and $\delta\Phi_{0} = \Phi
(t_{0}) \,- \,1$ for $\Lambda = 0$. Further, because of the matching
conditions at $t_{d}$, relations (\ref{sol. psi J-F cosmological rho
+ lambda approx 1}) and (\ref{sol. eq Phi J-F cosmological rho +
lambda approx 1}) are respectively subject to the constraints $$
\delta\psi_{\Lambda} = \frac{1}{\sqrt{\lambda}} \,\frac{4}{9}
\,{\beta}_{\psi}\,v \,[ \,1 \,- \,\frac{1}{2} \,\ln{\lambda} \,- \,
\,x_{d} \,\coth{x_{d}} $$ \begin{equation} \label{constraint at
matter-radiation decoupling epoch psi bis} + \,\ln(\sinh{x_{d}}) \, ]
\tanh{x_{d}} \end{equation} and $$\delta\Phi_{\Lambda} = \,-
\,\frac{4}{9} \,{\beta}_{\psi}\,v \,\delta\psi_{\Lambda}
\,\sqrt{\lambda} \,- \,\frac{1}{\sqrt{\lambda}} \,\frac{4}{9}
\,{\beta}_{\Phi}\,v \,[ \,1 \,- \,\frac{1}{2} \,\ln{\lambda} $$ $$-
\, \,x_{d} \,\coth{x_{d}} \,+ \,\ln(\sinh{x_{d}}) \, ] \tanh{x_{d}}
$$ $$ \,- \,\frac{1}{4} \,\delta\psi_{\Lambda}^{2} \,[ \,1 \,-
\,2\sqrt{\lambda}\,\sinh^{-1}{2x_{d}} \,] \,+\,\frac{2}{9}
\,{\beta}_{\psi}\,v \,\delta\psi_{\Lambda} \,[ \,\frac{1}{2} $$ $$+
\,\frac{1}{2} \,x_{d} \,( \,\tanh{x_{d}} \,+ \,\sinh{2x_{d}} \,) \,-
\,\ln{(\sinh{x_{d}})} $$ \begin{equation} \label{constraint at
matter-radiation decoupling epoch Phi bis} \,+
\,\frac{x_{d}}{\sinh{2x_{d}}} \,- \,\frac{1}{2} \,\sinh^{2}{x_{d}}
\,- \,\frac{1 \,+ \,\ln(\sinh{x_{d}})}{\cosh{x_{d}}} \,],
\end{equation} where we have set $x_{d} = \frac{3}{2}
\,\sqrt{\lambda} \,H_{0}\,t_{d}$, that is $$ x_{d} = \ln (
\,\,\sqrt{\frac{\lambda}{{\Omega}_{m}}} \,( \,1 \,+ \,z_{d} \,
)^{-3/2} $$ \begin{equation} \label{xd vs z} + \,\sqrt{1 \,+
\,\frac{\lambda}{{\Omega}_{m} \,( \,1 \,+ \,z_{d} \, )^{-3}} \,\,)}.
\end{equation} Clearly, because of the constraints (\ref{constraint
at matter-radiation decoupling epoch psi}), (\ref{constraint at
matter-radiation decoupling epoch Phi}), (\ref{constraint at
matter-radiation decoupling epoch psi bis}) and (\ref{constraint at
matter-radiation decoupling epoch Phi bis}), only two parameters (the
coupling constants ${\beta}_{\Phi}\,v$ and ${\beta}_{\psi}\,v$) are
left free to fit the data.

\item {\bf The future universe (cosmological constant era)}

In the future of the universe, the matter density and pressure become
quite negligible with respect to the $\Lambda$ term: $P = \rho=0$,
still no spatial curvature. Putting $a \propto \exp{(-
\sqrt{\Lambda/3} \,\,ct)}$ assuming $\Lambda > 0$, we obtain
\begin{equation} \label{eq psi J-F cosmological rho approx 1 Lambda
era} \ddot{\psi} \,+ \,c \,\,\sqrt{3\Lambda} \,\,\dot{\psi} = 0,
\end{equation} \begin{equation} \label{eq Phi J-F cosmological rho
approx 1 Lambda era} \ddot{\Phi} \,+ \,c \,\,\sqrt{3\Lambda}
\,\,\dot{\Phi} = - \,\frac{1}{2} \,{\dot{\psi}}^{2}. \end{equation}
The solutions of equations (\ref{eq psi J-F cosmological rho approx 1
Lambda era}) and (\ref{eq Phi J-F cosmological rho approx 1 Lambda
era}) are respectively of the same form as (\ref{sol. psi J-F
cosmological rho approx 1 radiation era}) and (\ref{sol. eq Phi J-F
cosmological rho approx 1 radiation era}). Hence, each scalar field
tends to a constant equal to its VEV, during the cosmological
constant era. As a consequence, here again the effective fine
structure constant will approach asymptotically the true fine
structure constant.

\end{enumerate}
\section{Comparison with the observational data}

We compare our prediction (\ref{alphaeff}) of the time variation of
$\alpha _{eff}$, for the two different cosmological models, with the
observational data. A quasar of redshift $z =
\frac{a(t_{0})}{a(t_{e})} \,- \,1$ emits photons at $t_{e}$, that we
receive at $t_{0}$ on Earth. Defining ${\alpha}_{z} = {\alpha}_{eff}
(t_{e})$, ${\alpha}_{0} = {\alpha}_{eff} (t_{0})$ and $\Delta
{\alpha}_{z} = {\alpha}_{z} \,- \,{\alpha}_{0}$, a least-squares fit
to the observational data (figure 1 of \cite{Webb}) gives,

\begin{itemize}
\item for $\lambda = 0$: ${\beta}_{\Phi}\,v \simeq 8.014~10^{-7}$ and
$\mid {\beta}_{\psi}\,v \mid \,\simeq 0.0793$, with ${\chi}^{2} =
0.948$ per degree of freedom (dof).\\

\item for $\lambda = 0.7$: ${\beta}_{\Phi}\,v \simeq 3.393~10^{-6}$ and
$\mid {\beta}_{\psi}\,v \mid \,\simeq 0.0317$, with ${\chi}^{2} =
0.779$ per dof. \end{itemize}
Figure 1 shows, in the same plot, the   observational results 
\cite{Webb} and our theoretical prediction
(\ref{alphaeff}) of $\Delta {\alpha}_{z}/{\alpha}_{0}$ versus the
redshift, for our two models  $\lambda = 0$ and $\lambda = 0.7$
(assuming  no spatial curvature).\\

The consistency of the recent observation from the distant quasars
absorption line spectra with the constraints from the Oklo uranium
deposit have been discussed in \cite{Webb}. At the corresponding
redshift, our best fits imply (assuming a spatially flat universe):\\
$( \,{\alpha}_{Oklo} \,- \,{\alpha}_{0} \,)/{\alpha}_{0} \simeq
-\,0.41 \,10^{-7}$ for $\lambda = 0$\\  $( \,{\alpha}_{Oklo} \,-
\,{\alpha}_{0} \,)/{\alpha}_{0} \simeq -\,1.9 \,10^{-7}$ for $\lambda
= 0.7$.\\
Including the Oklo point in the fit modify the $\chi ^{2}$
as indicated in the figure caption, and we conclude that our two best
fits are consistent with the Oklo bounds. This gives an averaged
decreasing rate approximately equal to $-\,10^{-17}$ per year,
consistent with the recent analyses \cite{Oklo bounds} on account of
the remark made in \cite{Webb} and in \cite{Fujii} for the case of a
non-linear time-evolution in $\Delta{\alpha}_{z}/{\alpha}_{0}$. Note
that this implies also a non trivial cosmic evolution for G$_{eff}$
which yields ${\dot{G}}_{eff}/G_{eff} \simeq -\,1.6~10^{-17}$ per
year at present, consistent with all the current bounds.\\

\begin{figure}[h] \begin{center} \begin{tabular}{c}
\epsfig{file=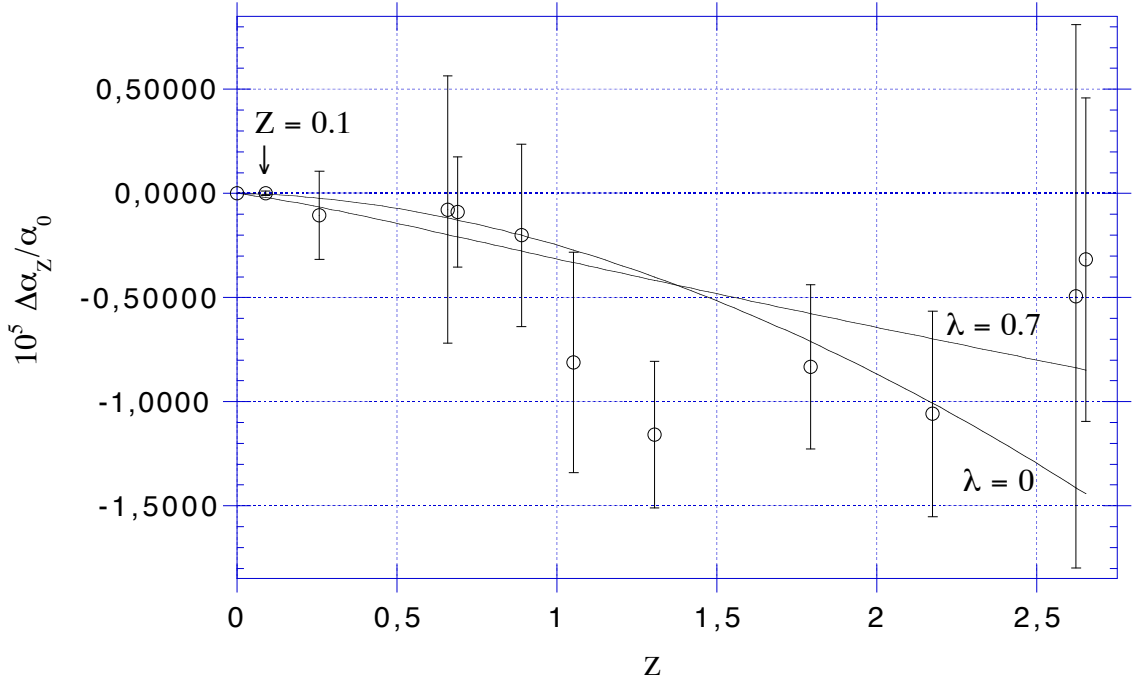,width=15.0cm} \end{tabular}
\end{center} \caption{Observed data and predicted curve $\Delta
{\alpha}_{z}/{\alpha}_{0}$ versus the redshift. The fits correspond
respectively to ${\chi}^{2} = 0.948$ ($\lambda = 0$) and ${\chi}^{2}
= 0.779$ ($\lambda = 0.7$) per dof, which seems to favor the late
time $\lambda$ dominated cosmology. Including the Oklo bounds ($z =
0.1$) in the data set would yield respectively ${\chi}^{2} = 0.882$
($\lambda = 0$) and ${\chi}^{2} = 1.094$ ($\lambda = 0.7$) per dof.}
\label{fig:cosmological alpha} \end{figure}

\section{Discussion and conclusion}

In the radiation dominated era, energy is present in the form of
radiation only, so that  the source terms for the scalar fields (equ.
\ref{eq psi J-F cosmological rho approx 1 radiation era} and \ref{eq
Phi J-F cosmological rho approx 1 radiation era}) cancel. Therefore,
both $\psi$ and $\Phi$ remain close to their respective VEV's. As a 
consequence, the effective fine
structure constant ${\alpha}_{eff}$ remains practically constant and
close to the true fine structure constant $\alpha$. It follows both
for BBN and at $z = 1000$ (the epoch of matter-radiation decoupling),
that $\Delta{\alpha}_{z}/{\alpha}_{0} \simeq -\,1.5~10^{-7}$, much below
the present observational bound (model dependent) which gives $\mid
\Delta{\alpha}_{z}/{\alpha}_{0} \mid \,< 10^{-4} \,- \,10^{-2}$ (see
\cite{BBN and CMB bounds}).\\

At the onset of the matter-$\Lambda$ dominated era, the scalar fields
will continuously start to vary ($\Phi$ increases), though at a
lesser extent than the density the ordinary matter. The Hubble
friction term introduces a relaxation time $\tau$ of the order
$1/3H$. This provides a na\-tu\-ral and sufficient way of driving back
${\alpha}_{eff}$ ($\simeq \alpha /{\Phi}^{3}$) to its constant value,
$\alpha$, after a lapse of time of a few $\tau$. As an estimate, let
us consider the Einstein-de Sitter cosmology: after $2\tau =
2/3H_{0} = t_{0}$, one gets $\mid {\alpha}_{eff} \, -\,\alpha
\mid_{max} \,\simeq e^{-2} \,\mid {\alpha}_{eff} \, -\,\alpha \mid_{t
= t_{0}}$. Since ${\alpha}_{eff} \simeq \alpha$ during the radiation
dominated era, it follows $\mid \Delta {\alpha}_{eff}/{\alpha}_{0}
\mid_{max} \,\simeq 0.135 \,\mid \Delta {\alpha}_{eff}/{\alpha}_{0}
\mid_{BBN}$. Hence, $\mid \Delta {\alpha}_{eff}/{\alpha}_{0}
\mid_{max} \,< 1.35~10^{-5} \,- \,1.35~10^{-3}$ as observed, on
account of the bound on the variation of the fine structure constant
at BBN.\\

Variations of the effective weak and strong coupling constants are
also expected in the higher dimensional theories candidates for
unification. The properties of the fundamental interactions are
connected to the topological properties of the compactified
extradimensions. Such theories involve more than one extradimension
in order to encompass all of the gauge groups of the standard model
of particle physics. In this framework, the effective constants of
the gauge fields would be expressed as functions of additional
internal fields ${\Phi}_{1}, ..., {\Phi}_{n}$. The effective
electromagnetic (fine structure constant), weak and strong coupling
constants would be written, respectively, as ${\alpha}_{eff} = \alpha
\,F_{1}({\Phi}_{1}, ..., {\Phi}_{n})$, ${\alpha}_{weff} =
{\alpha}_{w} \,F_{2}({\Phi}_{1}, ..., {\Phi}_{n})$ and
${\alpha}_{seff} = {\alpha}_{s} \,F_{3}({\Phi}_{1}, ...,
{\Phi}_{n})$. The functions $F_{1}$ and $F_{2}$ are related to each
other, because of the electroweak unification; and to $F_{3}$, if an
unification scheme is already present. Hence, we expect, at any given
time scale (dropping the $eff$ indexes for clarity):
$\frac{{\dot\alpha}_{w}}{{\alpha}_{w} }= \frac{\,\partial
\ln{F_{2}}}{\partial \ln{F_{1}}} \, \,\frac{{\dot\alpha}}{{\alpha}}$
and $\frac{{\dot\alpha}_{s}}{{\alpha}_{s}} = \frac{\,\partial
\ln{F_{3}}}{\partial \ln{F_{1}}} \, \,\frac{{\dot\alpha}}{{\alpha}} =
\frac{\,\partial \ln{F_{3}}}{\partial \ln{F_{2}}} \,
\,\frac{\,\partial \ln{F_{2}}}{\partial \ln{F_{1}}} \,
\,\frac{{\dot\alpha} }{{\alpha} }$. Expecting the ratios $\mid\frac{
\partial \ln{F_{2}}}{\partial \ln{F_{1}} }\mid$ and $\mid\frac{
\partial \ln{F_{3}}}{\partial \ln{F_{2}}} \mid$ of the order unity,
the three rates $\mid \frac{{\dot\alpha}_{w}}{{\alpha}_{w}} \mid$,
$\mid\frac{ {\dot\alpha}_{s}}{{\alpha}_{s}} \mid$ and $\mid
\frac{{\dot\alpha}}{{\alpha}} \mid$ should be comparable, both at BBN
and at the epoch of the Oklo phenomenon (see \cite{Uzan}).\\

We conclude that our modified Kaluza-Klein type action provides a
good effective description of interactions at low energy. The
instability problem \cite{Mbelek} is cured by the introduction of an
additional external bulk scalar field minimally coupled to gravity.
It accounts naturally for a cosmological time variation of $\alpha$,
in agreement with recent data. It also reconcile the discordant
laboratory measurements of $G$, by interpreting their differences as
due to a coupling with the dipolar magnetic field of the Earth
\cite{paper I}.\\

Compared to other explanations, our assumption appears as an
economical extension of general relativity. Like in the
Barrow-Sandvik-Magueijo (BSM) work \cite{Barrow}, the variation of
the effective fine structure constant is related to the coupling of a
scalar field to the Maxwell invariant
$F^{\alpha\beta}\,F_{\alpha\beta}$. Likewise, the effective fine
structure constant remains constant during the radiation era.
However, in contrast to BSM who predicts the suppression of the
changes in ${\alpha}_{eff}$ in the $\Lambda$ dominated era, in our
model the effective constants start to vary at the onset of the
matter-cosmological constant dominated era. Further, BSM find that
the product $G_{eff} \,{\alpha}_{eff}$ should approach an asymptotic
constant, whereas in this paper it is the quotient
$G_{eff}^{3}/{\alpha}_{eff}$ that remains approximately constant
during the cosmic evolution, since $G_{eff}$ and ${\alpha}_{eff}$
vary respectively as ${\Phi}^{-1}$ and approximately ${\Phi}^{-3}$.
Above all, the velocity of light in vacuum remains constant, as well
as the electric charges. Moreover, our model derives from a very
simple geometrical hypothesis. The first scalar field $\Phi$ (the
fifteen degree of freedom of the metric) is purely geometric, and
thus part of gravity in 5D. It affects directly the fine structure
constant and the gravitational constant. In contrary to $\psi$, it is
not minimally coupled to gravity (in 4D). Moreover, any kind of
matter (except $\psi$ itself) acts as a source for the external
scalar field $\psi$, and not only the EM field like in the BSM work.

The present model is limited and intends to be effective only. More
precise predictions would result from a fundamental theory, in the
same spirit.

\end{document}